\documentclass[prd,reprint,nofootinbib,showpacs,superscriptaddress]{revtex4-1}
\usepackage{graphicx} 
\usepackage{hyperref}
\usepackage{amsfonts}
\usepackage{amsmath,amssymb}
\usepackage{bm} 
\usepackage{color}
\usepackage{epstopdf} 
\usepackage{epsfig}
\usepackage{subfig} 
\usepackage{float}

{\rm }

\def\be{\begin{equation}}
 \def\ee{\end{equation}}
 \def\bea{\begin{eqnarray}}
 \def\eea{\end{eqnarray}}
 \def\bes{\begin{eqnarray}}
 \def\ees{\end{eqnarray}}
 \def\bi{\begin{itemize}}
 \def\ei{\end{itemize}} 

\def\2{\frac{1}{2}}
\def\4{\frac{1}{4}}

\begin{document}

\title{Generating the local oscillator ``locally'' in continuous-variable quantum key distribution based on coherent detection}

\author{Bing Qi}
\email{qib1@ornl.gov}
\affiliation{Quantum Information Science Group, Computational Sciences and Engineering Division,
Oak Ridge National Laboratory, Oak Ridge, TN 37831-6418, USA}
\affiliation{Department of Physics and Astronomy, The
University of Tennessee, Knoxville, TN 37996 - 1200, USA
}

\author{Pavel Lougovski}
\affiliation{Quantum Information Science Group, Computational Sciences and Engineering Division,
Oak Ridge National Laboratory, Oak Ridge, TN 37831-6418, USA}

\author{Raphael Pooser}
\affiliation{Quantum Information Science Group, Computational Sciences and Engineering Division,
Oak Ridge National Laboratory, Oak Ridge, TN 37831-6418, USA}
\affiliation{Department of Physics and Astronomy, The
University of Tennessee, Knoxville, TN 37996 - 1200, USA
}

\author{Warren Grice}
\affiliation{Quantum Information Science Group, Computational Sciences and Engineering Division,
Oak Ridge National Laboratory, Oak Ridge, TN 37831-6418, USA}

\author{Miljko Bobrek}
\affiliation{RF, Communications, and Intelligent Systems Group, Electrical and Electronics Systems Research Division, Oak Ridge National Laboratory, Oak Ridge, TN 37831-6006, USA}

\date{\today}

\begin{abstract}

Continuous-variable quantum key distribution (CV-QKD) protocols based on coherent detection have been studied extensively in both theory and experiment. In all the existing implementations of CV-QKD, both the quantum signal and the local oscillator (LO) are generated from the same laser and propagate through the insecure quantum channel. This arrangement may open security loopholes and limit the potential applications of CV-QKD. In this paper, we propose and demonstrate a pilot-aided feedforward data recovery scheme which enables reliable coherent detection using a ``locally'' generated LO. Using two independent commercial laser sources and a spool of $25$ km optical fiber, we construct a coherent communication system. The variance of the phase noise introduced by the proposed scheme is measured to be $0.04$ ($rad^2$), which is small enough to enable secure key distribution. This technology also opens the door for other quantum communication protocols, such as the recently proposed measurement-device-independent (MDI) CV-QKD where independent light sources are employed by different users.  

\end{abstract}

\maketitle

\section{Introduction}
\label{sec:1}

Quantum key distribution (QKD) allows two authenticated parties, normally referred to as Alice and Bob, to generate a secure key through an insecure quantum channel controlled by an eavesdropper, Eve \cite{BB84,E91,Gisin02,Scarani09,Lo14}. Based on fundamental laws in quantum mechanics, idealized QKD protocols have been proved to be unconditionally secure against adversaries with unlimited computing power and technological capabilities \cite{Mayers01,Lo99,Shor00}.

Both discrete-variable (DV) QKD protocols based on single photon detection \cite{BB84,E91} and continuous-variable (CV) QKD protocols based on coherent detection \cite{Ralph99,Hillery00,GMCS} have been demonstrated as viable solutions in practice. One well-known CV-QKD protocol is the Gaussian-modulated coherent state (GMCS) protocol \cite{GMCS}, which has been demonstrated through an $80$km optical fiber link recently \cite{Jouguet13}. One important advantage of the GMCS QKD is its robustness against incoherent background noise. The strong local oscillator (LO) employed in coherent detection also acts as a natural and extremely selective filter, which can suppress noise photons effectively. This intrinsic filtering function makes CV-QKD an appealing solution for secure key distribution over a noisy channel, such as a lit fiber in a conventional fiber optic network \cite{Qi10,Jouguet132,Kumar15} or a free-space optical link \cite{Heim14}.

However, all existing implementations of CV-QKD based on coherent detection contain a serious weakness: to reduce the phase noise, both the signal and the LO are generated from the same laser and propagate through the insecure quantum channel \cite{GMCS,Jouguet13,Qi07,Heim14} \footnote{Note in \cite{Heim14}, continuous polarization states (that contain the signal encoding as well as a LO in the same spatial mode) are prepared and sent over a free-space link in a polarization multiplexed setting. Such a configuration can automatically offer a high interferometric visibility.}. This arrangement has several limitations. First of all, it allows Eve to access both the quantum signal and the LO. Eve may launch sophisticated attacks by manipulating the LO, as demonstrated in recent studies \cite{Ma13,Huang13,Jouguet133,Huang14}. Second, sending a strong LO through a lossy channel can drastically reduce the efficiency of QKD in certain applications. For example, to achieve a shot-noise limited coherent detection, the required photon number in the LO is typically above $10^8$ photons per pulse at the receiver's end \cite{GMCS,Jouguet13,Qi07}. With a $1$ GHz pulse repetition rate and a channel loss of $20$ dB, the required LO power at the input of the quantum channel is about $1.2$ W (at $1550$ nm). If optical fiber is used as the quantum channel, noise photons generated by the strong LO inside the optical fiber may significantly reduce QKD efficiency and multiplexing capacity. Third, the LO is typically $7$ or $8$ orders of magnitude brighter than the quantum signal, complicated multiplexing and demultiplexing schemes are required to effectively separate the LO from the quantum signal at the receiver's end \footnote{Note, the second and third problems discussed above might be solved by sending a weak LO from Alice and applying optical amplification at Bob's side.}.

In brief, in CV-QKD, it is desirable to generate the LO ``locally'' using an independent laser source at the receiver's end. Unfortunately, such a scheme has never been implemented in practice. The main challenge is how to effectively establish a reliable phase reference between Alice and Bob. While various techniques, such as feedforward carrier recovery \cite{Ip07}, optical phase-locked loops \cite{Park12}, and optical injection phase-lock loop \cite{Fice11}, have been developed in classical coherent communication, these techniques are not suitable in QKD where the quantum signal is extremely weak and the tolerable phase noise is low. Furthermore, to prevent Eve from manipulating the LO, the LO laser should be isolated from outside both optically and electrically.

In this paper, we solve the above long outstanding problem by proposing and demonstrating a pilot-aided feedforward data recovery scheme, which enables reliable coherent detection using a ``locally'' generated LO. This scheme is built upon the observation that in the GMCS QKD, Bob does not need to perform the measurement in the ``correct basis''. In fact, Bob can perform the measurement in an arbitrarily rotated basis as along as the basis information (the phase reference) is available afterwards. With this post-measurement basis information, either Alice or Bob can rotate data at hand and generate correlated data with the other. We demonstrate the above scheme in a coherent communication system constructed by a spool of $25$ km optical fiber and two independent commercial laser sources operated at free-running mode. The observed phase-noise variance is $0.04$ ($rad^2$), which is small enough to enable secure key distribution. This technology also opens the door for other novel quantum communication protocols, such as the measurement-device-independent (MDI) CV-QKD protocol \cite{Pirandola13, Li14, Ma14} where independent light sources are employed by different users.  

This paper is organized as follows: in Section \ref{sec:2}, we conduct a theoretical analysis of the proposed scheme. In Section \ref{sec:3}, we present the details of proof-of-principle experiments. We conclude this paper with a discussion in Section \ref{sec:4}.
 
\section{Theoretical analysis}
\label{sec:2}

In GMCS QKD, Alice draws two random numbers $X_A$ and $P_A$ from a set of Gaussian random numbers (with a mean of zero and a variance of $V_AN_0$), prepares a coherent state $|X_A+iP_A\rangle$ accordingly, and sends it to Bob. Here $N_0$ = 1/4 denotes the shot-noise variance. At Bob's end, he can perform either optical homodyne detection or optical heterodyne detection.

In GMCS QKD protocol based on homodyne detection \cite{GMCS}, Bob randomly chooses to measure either the amplitude quadrature (X) or phase quadrature (P) of the incoming signal. Later on, he announces which quadrature he measures for each incoming signal through an authenticated public channel, and Alice only keeps the corresponding data. In GMCS QKD based on heterodyne detection \cite{Weedbrook04}, Bob first splits the incoming signal into two with a 50:50 beam splitter. He then measures X at one output port and P at the other. In this case, Alice keeps all her quadrature data.

After the quantum transmission stage, Alice shares a set of correlated Gaussian random variables (called the``raw key'') with Bob. Alice and Bob compare a random sample of the raw key through an authenticated classical channel to estimate the transmittance and excess noise of the quantum channel. If the observed excess noise is small enough, they can further work out a secure key.

In the above description, we have implicitly assumed that Alice and Bob share a phase reference, so Bob can perform the required quadrature measurement. If the LO is generated for an independent laser source, how can Alice and Bob establish a phase reference in this case?

In this section, we present a pilot-aided phase estimation scheme which allows Alice and Bob to measure the phase relation between two independent lasers in real time. Using this phase information either Alice or Bob can rotate the data at hand in the post-processing stage (``quadrature remapping'') and establish correlation with the other. In principle, our scheme can be applied to both CV-QKD with homodyne detection and the one with heterodyne detection. In this paper, we focus on the case of heterodyne detection. For an independent and related work, see \cite{Soh15}.

\subsection{CV-QKD using quadrature remapping scheme}

In a phase coding DV-QKD protocol, it is also crucial to control the phase between a signal pulse and a reference pulse when performing interferometric measurement. In fact, a DV-QKD protocol using a strong phase-reference pulse has been proposed in \cite{Koashi04}. In this scheme, Alice sends Bob a quantum signal together with a strong phase reference pulse generated from the same laser. At Bob's side, he interferes the strong phase reference pulse with a sampling beam from his LO laser to determine the phase difference between the two lasers, corrects this phase difference by introducing a phase shift to his LO laser, and then performs an interfeometric measurement on the quantum signal using the phase-corrected LO pulse. However, the above scheme has not been demonstrated yet, possibly due to the following reasons: first, the phase difference between two remote independent lasers is expected to fluctuate rapidly, this makes real-time phase feedback control very challenging. Secondly, different types of detector are required for phase measurement and quantum signal detection, this increases the complexity of the overall system. As we will show below, the above two challenges can be overcome in a CV-QKD protocol.

Suppose in a CV-QKD system based on heterodyne detection, both the signal laser and the LO laser are operated in free-running mode. Without loss of generality, for each transmission, we can choose the phase of the signal laser as the phase reference ($\phi_S=0$). When Bob performs conjugated homodyne detection, the phase $\phi$ of his LO laser can be treated as a random variable. Bob's measurement results $(X_B,P_B)$ are given by (after scaling with the channel transmittance)
\bes\label{eq1} X_B &=& X_A cos\phi+P_A sin\phi+N_X \nonumber\\
P_B &=& -X_A sin\phi+P_A cos\phi+N_P \ees
where $N_X$ and $N_P$ are assumed to be i.i.d. Gaussian noises with zero mean. 

If Alice and Bob can determine $\phi$ after Bob has performed his measurement, one of them (for example, Bob) can use this post-measurement phase information to correct his data by performing the following rotation
\bes\label{eq2} X^{'}_B &=& X_B cos\phi-P_B sin\phi \nonumber\\
P^{'}_B &=& X_B sin\phi+P_B cos\phi \ees

From equations (1) and (2), it is easy to show
\bes\label{eq3} X^{'}_B &=& X_A+N^{'}_X \nonumber\\
P^{'}_B &=& P_A+N^{'}_P \ees
where the noise terms in the rotated data are given by
\bes\label{eq4} N^{'}_X &=& N_X cos\phi-N_P sin\phi \nonumber\\
N^{'}_P &=& N_X sin\phi+N_P cos\phi \ees

Given $N_X$ and $N_P$ are i.i.d. Gaussian noises, it is easy to see that $N^{'}_X$ and $N^{'}_P$ are also independent Gaussian noises with the same variance as $N_X$ and $N_P$. This suggests that the rotation process will not introduce additional noise if the phase $\phi$ can be determined precisely. 

The above ``quadrature remapping'' scheme allows Alice and Bob to establish correlated data without using a complicated phase feedback control system, thus removing the first challenge listed at the beginning of this section. Next, we will present a scheme which allows Alice and Bob to determine $\phi$ under \textbf{realistic} scenarios using the \textbf{same} detector for quantum signal detection, thus removing the second challenge listed above.

\subsection{Pilot-aided phase recovery scheme}

If the drift of phase $\phi$ is slow enough such that within a frame time of $\Delta T$ (within which the phase $\phi$ can be treated as a constant), many rounds of quantum transmission can be conducted, the following scheme can be applied to estimate the phase $\phi$. After the quantum transmission stage, for each frame, Alice can randomly choose a subset of the transmitted signals as calibration pulses and announce the encoded data through an authenticated channel. Using the corresponding measurement results at hand, Bob can estimate phase $\phi$ for this frame using equation (1). Since Alice's signals are at quantum level, each individual calibration pulse cannot provide a precise estimation of the phase $\phi$. However, by averaging the results acquired from a large number of calibration pulses, the phase noise can be reduced effectively. This scheme was first proposed and implemented in \cite{Qi07} to reduce the noise associated with the slow phase drift of a fiber interferometer in GMCS QKD.

Unfortunately, the above scheme is not practical when the quantum signals and the LOs are generated from independent laser sources. On one hand, the phase difference between two practical lasers fluctuates rapidly due to the laser frequency instability and the phase noise associated with the finite laser linewidth; on the other hand, the maximum transmission rate of CV-QKD is limited by the bandwidth of shot-noise limited optical coherent detector. As such, we cannot acquire an accurate estimation of $\phi$ by measuring quantum signals.

To solve the above problem, we proposed a pilot-aided feedforward data recovery scheme \cite{Qi13}. The basic idea is as follows. For each quantum transmission, Alice sends out both a quantum signal and a relatively strong phase reference pulse generated from the same laser. The quantum signal carries Alice's random numbers, as in the case of conventional CV-QKD. The reference pulse, on the other hand, is not modulated. These two pulses propagate through the same quantum channel to the measurement device, where Bob performs conjugate homodyne detection on both of them using LOs generated from the LO laser. Note, to avoid detector saturation, Bob can use a relatively weak LO to measure the reference pulse. 

The measurement results from the phase reference pulse $(X_R,P_R)$ can be used to determine $\phi$ using
\bes\label{eq5} \phi=-tan^{-1}\dfrac{P_R}{X_R}\ees
where the minus sign is due to the definition of phase reference. By using a relatively strong reference pulse, Bob can acquire an accurate estimation of $\phi$ and use this phase information to implement the quadrature remapping scheme.

In this paper, we study a simple implementation of the above scheme, where Alice sends out quantum signals and reference pulses alternately and periodically, as shown in Figure 1. We remark that if the drift of phase $\phi$ is slow enough compared with the transmission rate of QKD, it is possible to use fewer reference pulses to improve QKD efficiency.

\begin{figure}[t]
	\includegraphics[width=.4\textwidth]{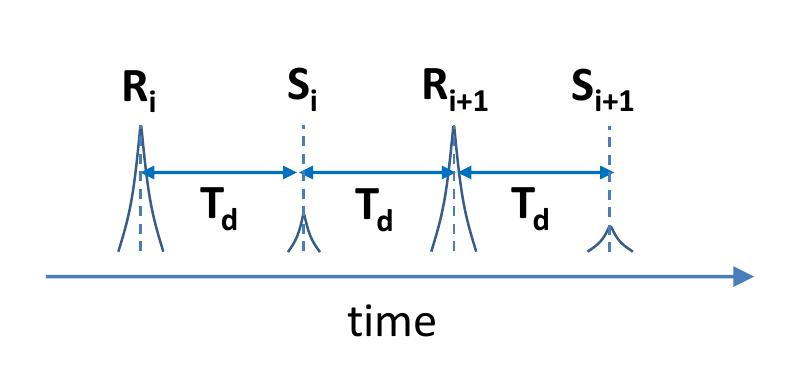}
	\captionsetup{justification=raggedright,
					singlelinecheck=false }
	\caption{Distribution of quantum signals ($S$) and reference pulses ($R$). } 
	\label{fig:1}
\end{figure}

In Fig.1, a quantum signal $S_i$ and the corresponding reference pulse $R_i$ are measured at different times with a time delay of $T_d$. If the frequency difference of the two lasers $(f_1-f_2)$ is a constant and can be precisely determined, we can estimate phase $\phi_{S,i}$ at the time when $S_i$ is measured from the phase measurement result of $R_i$ by simply adding a constant phase shift of $2\pi (f_1-f_2)T_d$. In practice, however, both lasers present slow frequency drift over time. Here, we use a simple scheme to estimate $\phi_{S,i}$.  Since the signal pulse $S_i$ is in the middle of two reference pulses $R_i$ and $R_{i+1}$, we can estimate $\phi_{S,i}$ from the phase measurement results on $R_i$ and $R_{i+1}$ as
\bes\label{eq6} \overline{\phi}_{S,i}=\dfrac{\phi_{R,i}+\phi_{R,i+1}}{2} \ees

Note the above equation can also be written as 
\bes\label{eq7} \overline{\phi}_{S,i}=\phi_{R,i}+2\pi \overline{f}_d T_d \ees
where $\overline{f}_d=\dfrac{\phi_{R,i+1}-\phi_{R,i}}{4\pi T_d}$ can be interpreted as the frequency difference of the two lasers within the short time interval between two adjacent reference pulses.

While similar to classical intradyne detection, a key difference in our scheme separates phase recovery of a quantum signal from that of a classical one. A phase reference cannot be recovered reliably from a quantum signal while it can in the classical case, meaning that the reference pulses here must be used to estimate that phase of the LO and quantum signal during the time window in which the quantum signal arrives. This places additional stringent requirements on relative laser noise compared to the classical case.

\subsection{Security analysis}

In this section, we will show that the existing security proofs of conventional CV-QKD \cite{Renner09, Leverrier09, Leverrier15} (built upon the assumption that Eve can only access the quantum signals) can be applied in our scheme directly.

First, the phase reference pulses are only used to provide (classical) phase information, they are not directly used in the coherent detection of the quantum signals. In fact, in our scheme Eve can never access the LO itself. Note, a standard assumption in CV-QKD is that Eve can have full knowledge of the phase reference used in quantum state preparation/coherent detection, so the reference pulses will not give Eve any additional information. Eve can certainly interfere with the phase recovery process by manipulating the phase reference pulses when they propagate through the quantum channel. This could result in an increased phase noise and the secure key rate will be reduced. This is one type of denial-of-service attack, which can be applied to any QKD protocols. From Eve's point of view, whatever can be achieved by manipulating the reference pulses can also be achieved by manipulating the quantum signals directly. In brief, sending phase reference pulses through the quantum channel will not cause any security problem.

Next, we will show the security of the CV-QKD protocol using quadrature remapping scheme is equivalent to that of the conventional CV-QKD protocol. To illustrate the essential ideas, it is convenient to represent the phase recovery scheme by a separate classical communication channel, which can be fully controlled by Eve. Fig.2 (a) is a schematic diagram of Bob's system in our new QKD scheme. In this picture, Bob performs a heterodyne measurement on the incoming quantum signal, then rotates his measurement results using the phase $\phi$ estimated through the classical communication channel. In \cite{Leverrier13}, the authors proved that a unitary phase rotation commutes with heterodyne detection. More specifically, Bob can either rotate the optical phase of the quantum signal first, then perform heterodyne detection, or he can perform heterodyne detection first, then rotate the classical measurement results in the post-processing stage. So the protocol shown in Fig.2 (a) is equivalent to the virtual QKD protocol shown in Fig.2 (b). Since the classical phase estimation channel can be controlled by Eve, we can move the phase rotation operator out of Bob's secure station and let Eve to have full control of it, as shown in Fig.2 (c). Note the QKD protocol shown in Fig.2 (c) is exactly the conventional CV-QKD based on heterodyne detection, where Eve is allowed to manipulate the quantum signals transmitted through the channel at her will. So, the security of our new QKD scheme is equivalent to that of the conventional CV-QKD protocol.

\begin{figure}[t]
	\includegraphics[width=.4\textwidth]{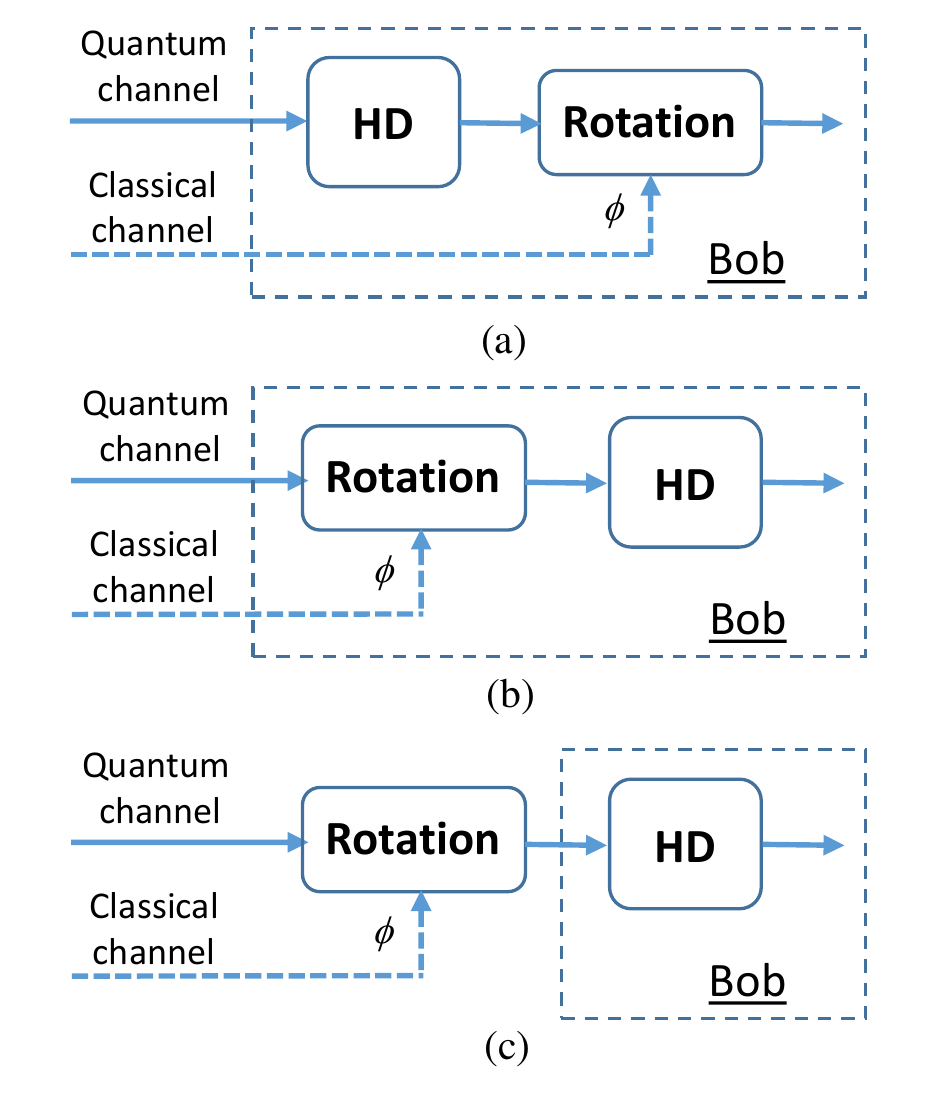}
	\captionsetup{justification=raggedright,
					singlelinecheck=false }
	\caption{Security models. HD--heterodyne detection. (a) CV-QKD protocol using quadrature remapping scheme; (b) A virtual QKD scheme equivalents to (a); (c) Conventional QKD scheme. } 
	\label{fig:2}
\end{figure}

While we do not need to develop a new security proof for the proposed QKD scheme, to achieve a high secure key rate, it is important to reduce the noise of the phase recovery process. From Eq.\ \eqref{eq2}, the uncertainty of $\phi$ will be translated into an excess noise in $(X^{'}_B,P^{'}_B)$ (after scaling with the channel transmittance) as
\bes\label{eq8} \varepsilon_\phi=V_A \sigma_\phi \ees
where $V_A$ is Alice's modulation variance, $\sigma_\phi$ is the noise variance in determining phase $\phi$. This extra noise $\varepsilon_\phi$ will reduce secure key generation rate. It is thus very important to minimize the phase noise $\sigma_\phi$.

In next section, we will study the performance of the proposed phase recovery scheme under realistic scenario.

\section{Proof of principle demonstration}
\label{sec:3}

\subsection{Noise model}

There are two major noise sources in determining phase $\phi$ using Eq.(6). The first one is the measurement noise when Bob tries to determine $\phi_{R,i}$ ($\phi_{R,i+1}$) of the reference pulses $R_i$ ($R_{i+1}$). This noise could be significant when the reference pulses become extremely weak thus the contribution of shot noise cannot be ignored. However, in practice, we can use a relative strong reference pulse to reduce the contribution of the shot noise. For example, if the average photon number of the reference pulse (at Bob's heterodyne detector) is $1000$, given the detection efficiency of the heterordyne detector is $50\%$, the phase noise variance due to the shot noise is about $0.001$, which is negligible in practice. In this paper, we simply ignore this noise contribution.

The second  noise source is the quantum phase noise of the laser, which originates from the amplified spontaneous emission. More specifically, even we know the phase of the reference pulse, we still cannot determine the phase of the signal pulse precisely since they are generated at different times. The spontaneous emitted photons generated within the above time interval contribute a fundamental phase noise. Since the laser phase noise cannot be reduced by simply increasing the amplitude of the reference pulse, it is the main noise source in our scheme.

Define the laser phase at time $t=0$ as $\theta_0$. The phase noise $\Delta\theta(t)$ quantifies the deviation of the laser phase at time $t$ from $\theta_0+2\pi f t$ (the phase expected from an ideal $sine$ wave), where $f$ is the central frequency of the laser. $\Delta\theta(t)$ can be modeled as a Gaussian random variable with a mean of zero and a variance of \cite{Yariv07}
\bes\label{eq9} \langle(\Delta\theta(t))^2\rangle=\frac{2t}{\tau_c}. \ees
Here $\tau_c$ is the coherence time of the laser. For a laser with lorentzian lineshape, $\tau_c$ is related to its linewidth $\Delta f$ by \cite{Yariv07}
\bes\label{eq10} \tau_c\simeq\frac{1}{\pi\Delta f}.\ees

As shown in Appendix A, given the phase noise of the signal laser and that of the LO laser are $\langle(\Delta\theta_S(t))^2\rangle$ and $\langle(\Delta\theta_L(t))^2\rangle$ respectively, the noise variance of our phase estimation scheme (Eq.(6)) is described by
\bes\label{eq11} \sigma_{\phi}=\dfrac{1}{2} \left\lbrace \langle(\Delta\theta_S(T_d))^2\rangle + \langle(\Delta\theta_L(T_d))^2\rangle  \right\rbrace, \ees
where $T_d$ is the time delay between the signal pulse and the reference pulse (see Fig.1). 

\subsection{Experimental setup}

We demonstrate the pilot-aided feedforward data recovery scheme using commercial off-the-shelf devices. The experimental setup is shown in Fig.3. Two commercial frequency-stabilized continuous wave (cw) lasers at Telecom wavelength (Clarity-NLL-1542-HP from Wavelength Reference) are employed as the signal and the LO laser. Both lasers are operated at free-running mode with no optical or electrical connections between them. The central frequency difference between the two lasers can stay within $10$ MHz without doing any feedback controls. A LiNbO3 waveguide intensity modulator (EOSpace) is used to generate $8$ ns laser pulses at a repetition rate of $50$ MHz. Since half of the laser pulses are used as phase references, the equivalent data transmission rate in our experiment is $25$ MHz. A LiNbO3 waveguide phase modulator (EOSpace) is used to modulate the phase of the signal pulses.

\begin{figure}[t]
	\includegraphics[width=.45\textwidth]{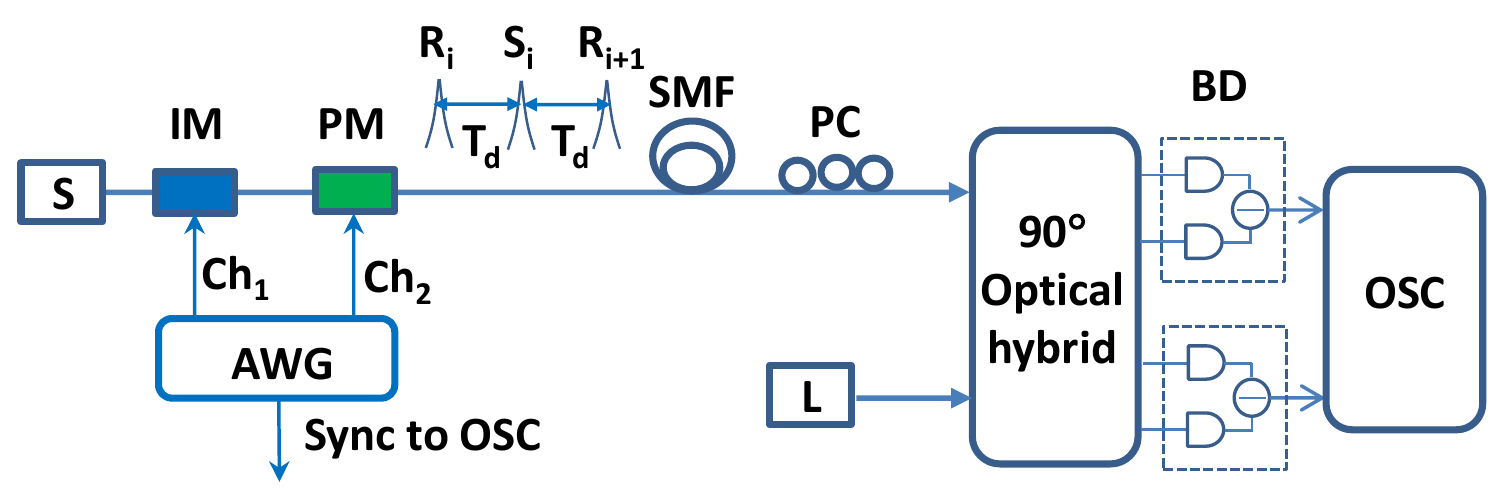}
	\captionsetup{justification=raggedright,
					singlelinecheck=false }
	\caption{Experimental setup. S-signal laser; L-LO laser; IM-optical intensity modulator; PM-optical phase modulator; AWG-arbitrary waveform generator; SMF-25km single mode fiber spool; PC- polarization controller; BD-balanced photodetector; OSC-oscilloscope. } 
	\label{fig:3}
\end{figure}

Both the signal pulses and the reference pulses propagate through a spool of 25km single mode fiber before arriving at the measurement device. A commercial $90^o$  optical hybrid (Optoplex) and two $350$ MHz balanced amplified photodetectors (Thorlabs) are employed to measure both X-quadrature and P-quadrature of the incoming pulses. The $90^o$ optical hybrid is a passive device featuring a compact design. No temperature control is required to stabilize its internal interferometers. The outputs of the two balanced photodetectors are sampled by a broadband oscilloscope at $1$ GHz sampling rate. For simplicity, the LO laser is operated at the cw mode. A waveform generator with a bandwidth of $120$ MHz provides the modulation signals to both the intensity and the phase modulator, and a synchronization signal to the oscilloscope.

\subsection{Experimental results}

To evaluate the effectiveness of the phase recovery scheme, we conduct a phase encoding coherent communication experiment using a binary pattern of ``01010101...'', where bit 0 is represented by no phase shift and bit 1 by phase shift of $1.65 rad$. The phase modulator shown in Fig.3 is used to encode binary phase information on the signal pulses. The amplitude of the signal pulse is the same as that of the reference pulse. At the receiver's end, the average photon number per pulse is about $10^5$, which is significantly lower than that of the LO used in a typical GMCS QKD experiment. Note, in this experiment, to determine the noise of the phase recovery scheme, strong signal pulses are employed to provide ``true'' values of the phases to be estimated. 

In total, $25000$ signal pulses and $25000$ reference pulses are transmitted. For each pulse received by Bob, its phase is calculated from the measured quadrature values $\left\lbrace X,P\right\rbrace $ using Eq.(5). The phase measurement results from the signal pulses $\left\lbrace \phi_{S,i}^{(raw)},  i=1,2,...25000\right\rbrace$ are shown in Fig.4(a) and (b). Due to the random phase change between the signal laser and the LO laser, the measured phases are randomly distributed within $[0,2\pi)$, regardless the encoded phase information.

From the phase measurement results of the reference pulses $\left\lbrace\phi_{R,i},  i=1,2,...25000\right\rbrace$, we recover a phase reference $\overline{\phi}_{S,i}$ for each signal pulse using Eq.(6), and correct the raw measurement results by 
\bes\label{eq12} \phi_{S,i}^{(cor)}=\phi_{S,i}^{(raw)}+\overline{\phi}_{S,i}\ees

The corrected phase measurement results $\left\lbrace \phi_{S,i}^{(cor)},  i=1,2,...25000\right\rbrace$ are shown in Fig.4(c) and (d). After the phase correction, the measurement results for bit 0 and bit 1 are clearly separated.

The variances of the residual phase noise (the difference between $\phi_{S,i}^{(cor)}$ and the phase information encoded by Alice) have been determined to be $0.040\pm0.001$ (for bit 0) and $0.039\pm0.001$ (for bit 1) respectively.

\begin{figure}[t]
	\includegraphics[width=.5\textwidth]{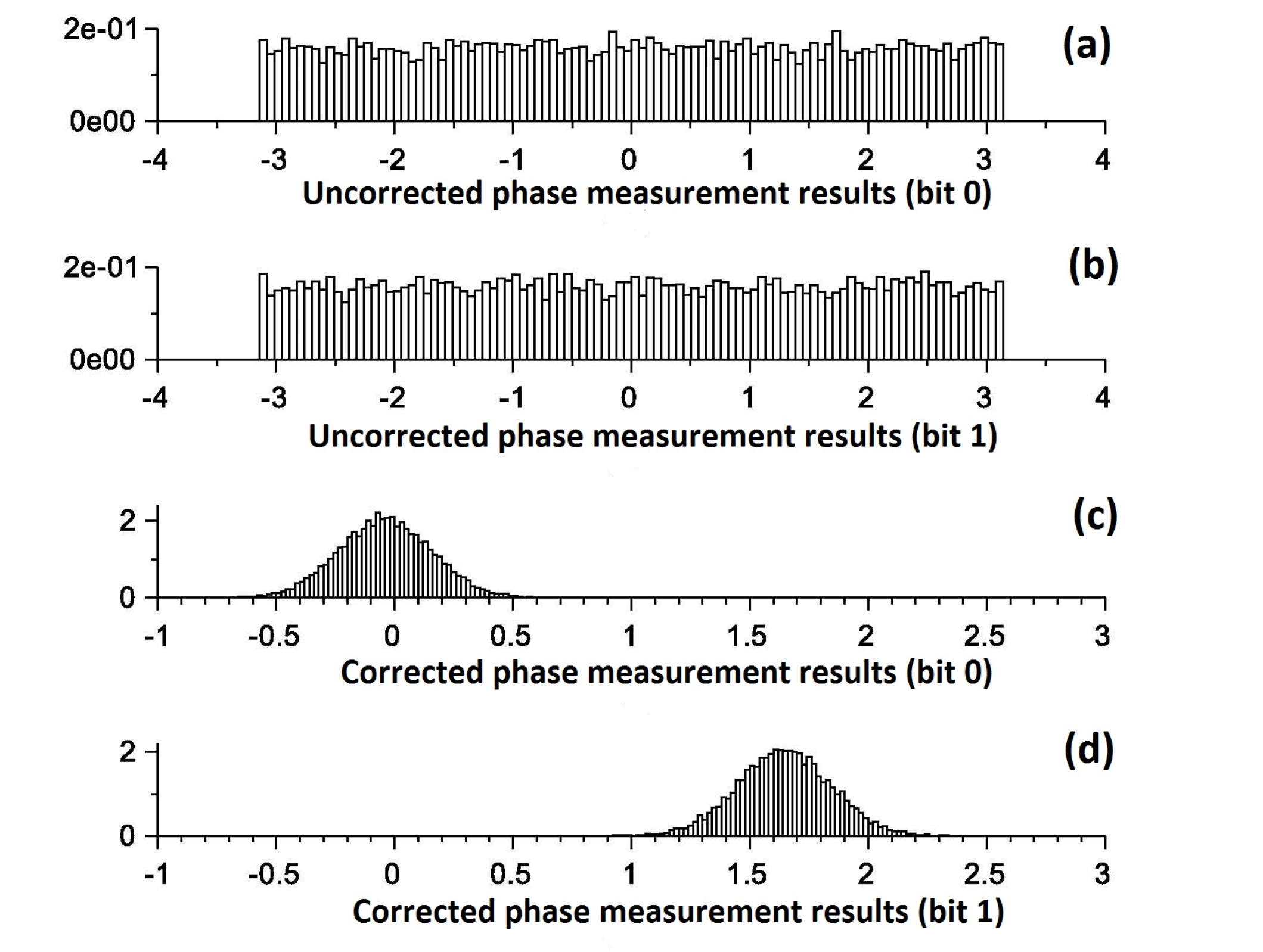}
	\captionsetup{justification=raggedright,
					singlelinecheck=false }
	\caption{Histograms of the phase measurement results. (a) The measurement results corresponding to bit 0 (before phase correction); (b) The measurement results corresponding to bit 1 (before phase correction); (c) The measurement results corresponding to bit 0 (after phase correction); (d) The measurement results corresponding to bit 1 (after phase correction). } 
	\label{fig:4}
\end{figure}

Note in the above experiment, relatively strong reference pulses have been employed. While this will not introduce any security problem, in practice, it may be more convenient to use weak reference pulses. We conduct experiments to  determine phase noise variance $\sigma_{\phi}$ using reference pulses with different average photon numbers $(10000, 1000, 100)$. The measured phase noise variance are $(0.039\pm0.001, 0.040\pm0.001, 0.054\pm0.001)$. These results show that the phase recovery scheme works well even with reference pulses containing only a thousand photons. 

As we have discussed in the previous section, the main noise source in our setup is laser phase noise associated with its finite linewidth. We conduct experiments to determine the phase noise of each laser. For $T_d=20ns$ (which corresponds to the 50MHz pulse repetition rate in the above experiments), the phase noise of the two lasers has been determined to be $0.035\pm0.001$ and $0.044\pm0.001$, see details in Appendix A. From Eq.(11), the expected noise of the phase recovery scheme is $\sigma_{\phi}=0.040\pm0.001$, which matches with the experimental results very well. To further reduce the noise $\sigma_{\phi}$, we can either use a smaller time delay $T_d$ (which is ultimately limited by detector bandwidth), or choose lasers with a narrower linewidth.

As another demonstration of the phase recovery scheme, we conduct an experiment by using the phase reference recovered from the reference pulses to remap quadrature values measured with weak quantum signals. In this experiment, no phase information is encoded on the signal pulses. The average photon number of each reference pulse at the receiver's end is about $1000$, while that of each signal pulse is $66$. Fig.5 shows the quadrature values $(X,P)$ of the signal pulses in phase-space (sample size is $24000$). The figure on the left shows the raw measurement results, where the phase randomly distributed in $[0,2\pi)$ as expected. The figure on the right shows the results after performing quadrature remapping. More specifically, we first recover a phase reference $\overline{\phi}_{S,i}$ for each signal pulse using Eq.(6), and then rotate the raw data using Eq.(2). The quadrature values have been scaled by taking into account the 3-dB loss due to heterodyne detection and the $50\%$ overall efficiency of the detection system. The noise variance in the X-quadrature (right figure) has been determined to be $1.83$ in shot noise units. This suggests the excess noise of the detector (including noise from the commercial balanced photo-detector and the oscilloscope) is about $0.83$ in shot-noise units. Note, due to the residual phase noise of the phase recovery scheme, the distribution shown in the right figure is not symmetric: the variance of P-quadrature ($\Delta_P$) is larger than that of X-quadrature ($\Delta_X$). The phase noise $\sigma_{\phi}$ in the above experiment can be estimated by $\sigma_{\phi}=(\Delta_P-\Delta_X)/X^2_{0}$, where $X_0$ is the mean value of X-quadrature. The experimental result is $(0.034\pm0.01)$, which is consistent with the noise variance estimated with strong signal pulses. This shows the proposed phase recovery scheme works well in both the classical and the quantum domain. Note the uncertainty in this measurement is higher than that in previous experiments, since we estimate a small quantity ($\sigma_{\phi}$) from the difference of two relatively large quantities ($\Delta_P$ and $\Delta_X$). 

\begin{figure}[t]
	\includegraphics[width=.5\textwidth]{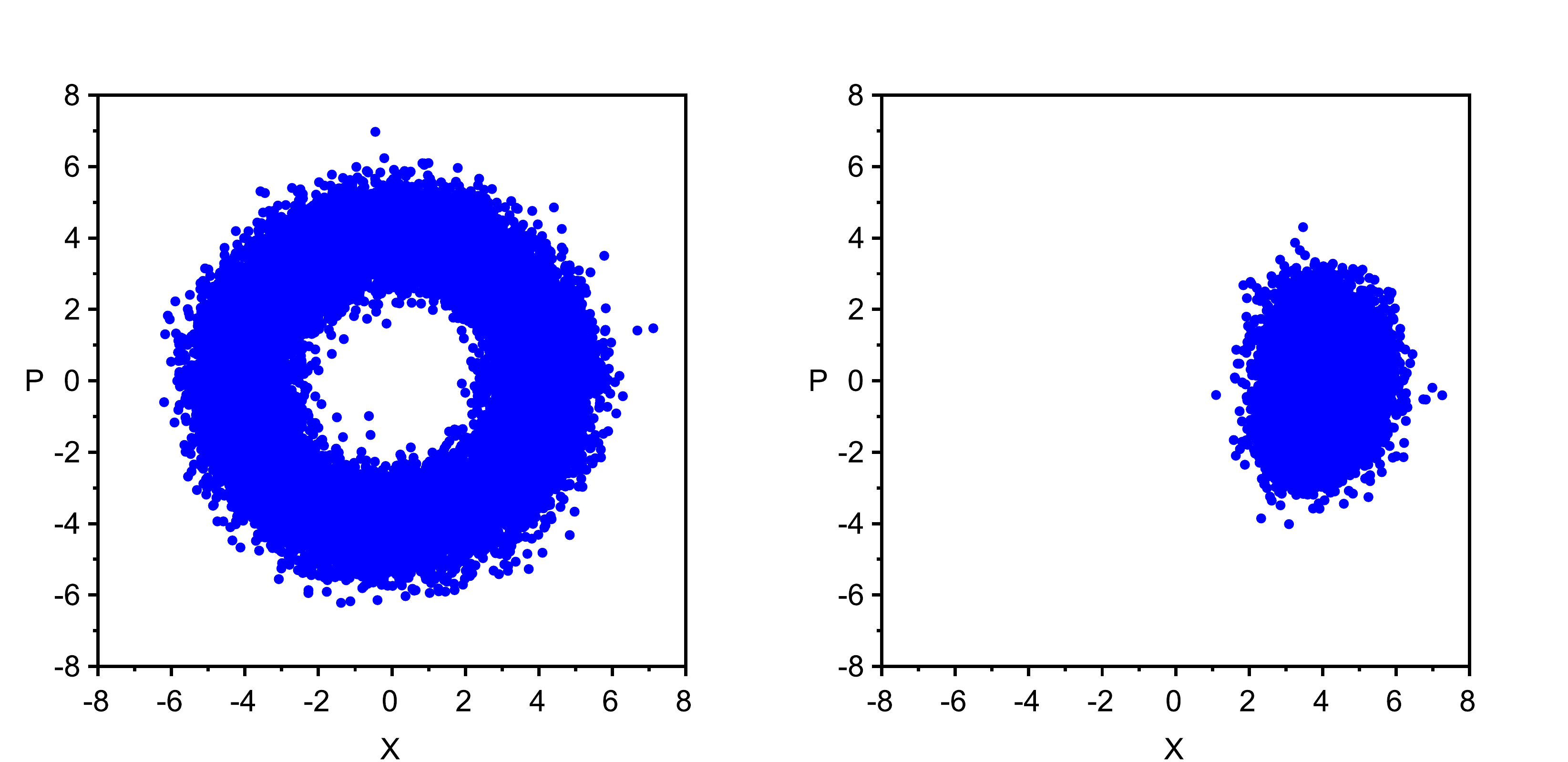}
	\captionsetup{justification=raggedright,
					singlelinecheck=false }
	\caption{The measured quadrature values in phase space. Left--before quadrature remapping; Right--after quadrature remapping. } 
	\label{fig:5}
\end{figure}

Given the noise of the phase recovery scheme, we can use Eq.(8) to determine the  additional excess noise contributed by this scheme and estimate the secure key rate using existing security proof of GMCS QKD. In appendix B, we present simulation results based on practical system parameters. Under the ``realistic'' model \cite{GMCS} where Eve cannot control the noise and loss of Bob's detector, secure key can be generated over a distance of 120km through telecom fiber. To estimate the finite data size effect, we also conduct simulations using the most recent composable security proof of CV-QKD \cite{Leverrier15}.

\section{Discussion}
\label{sec:4}

A long outstanding problem in CV QKD based on coherent detection is how to generate the LO ``locally''. In all the existing implementations of CV-QKD, both the quantum signal and the LO are generated from the same laser and propagate through the insecure quantum channel. This arrangement may open security loopholes and also limit the potential applications of CV-QKD.

In this paper, we solve the above problem by proposing and demonstrating a pilot-aided feedforward data recovery scheme which allows reliable coherent detection using a ``locally'' generated LO. This scheme also greatly simplifies the CV-QKD design by getting rid of the cumbersome unbalanced fiber interferometers and the associated phase stabilization system. Proof of principle experiments based on commercial off-the-shelf components show that the noise due to the proposed scheme is tolerable in CV-QKD. To further reduce the noise, laser sources with a smaller linewidth can be applied.

We remark that the measurement device employed in our experiment is essentially an intradyne detection scheme which has been applied in classical coherent communication for carrier phase recovery \cite{Derr92, Noe05}. It is thus convenient to name our new scheme as ``intradyne'' CV-QKD while the conventional scheme as ``self-homodyne'' CV-QKD \cite{Note}. However, there are several important differences between the classical and the quantum case. First, in classical communication, the signals are strong and the modulation scheme (such as BPSK and QPSK) is relatively simple. This allows carrier phase recovery from the signals directly. In GMCS QKD, the quantum signals are extremely weak and the modulation scheme is more complicated, the carrier phase cannot be recovered from the quantum signals reliably. Thus it is necessary to employ relatively strong reference pulses. Second, the transmission rate of a classical communication system can reach 100GHz, while the transmission rate of a state-of-the-art GMCS QKD system is below 100MHz. This places a more stringent requirement on laser phase noise in the quantum system. Third, a classical digital communication system can tolerate higher phase noise than the CV-QKD. In brief, it is much more challenging to recover carrier phase in quantum communication.

Although  a complete CV-QKD experiment using the proposed scheme is not presented in this paper, all the components required to implement such a system, including broadband shot-noise limited homodyne detectors \cite{Okubo08, Chi11, Kumar12}, have been well developed. In fact, the structure of the proposed QKD system is much simpler compared to the conventional scheme \cite{Jouguet13}.

We remark that a similar CV-QKD scheme has been independently proposed by Soh \textit{et al.} \cite{Soh15}. In \cite{Soh15}, Soh \textit{et al.} study the expected secure key rate of their protocol under a passive channel taking into account the effects of quantum noise on the reference pulse, and show in what limit the reference pulse scheme achieves the same performance as the standard scheme (where an LO is transmitted). They further conduct a proof-of-principle QKD experiment in the presence of strong phase noise between Alice's signal pulses and Bob's LO pulses generated from the same laser. In our study, we establish the security of the proposed QKD protocol by showing that it is equivalent to the conventional GMCS QKD protocol, thus the well-established security proof can be applied directly. Our proof-of-principle demonstration focuses on establishing a reliable phase reference between two independent lasers over a $25$km optical fiber link, a practical scenario that the proposed protocol is designed for. We expect our scheme will be widely adopted in CV-QKD. This technology also opens the door for other quantum communication protocols, such as the MDI-CV-QKD protocol.

We would like to thank Hoi-Kwong Lo and Paul Jouguet for very helpful discussions. This work was performed at Oak Ridge National Laboratory, operated by UT-Battelle for the U.S. Department of Energy under Contract No. DE-AC05-00OR22725. The authors acknowledge support from the laboratory directed research and development program.

\appendix

\section{Laser phase noise}

In this Appendix, we will first derive Eq.(11), which quantifies the contribution of laser phase noise to the noise variance of the phase recovery scheme. Then we will present details of experiments where the phase noise of each laser is measured.

For simplicity, we consider the case that the phases of two reference pulses  measured at time $t_0$ and $t_2$ are used to estimate the phase difference of the signal laser and the LO laser at time $t_1$, as shown in Fig.6. 

\begin{figure}[t]
	\includegraphics[width=.5\textwidth]{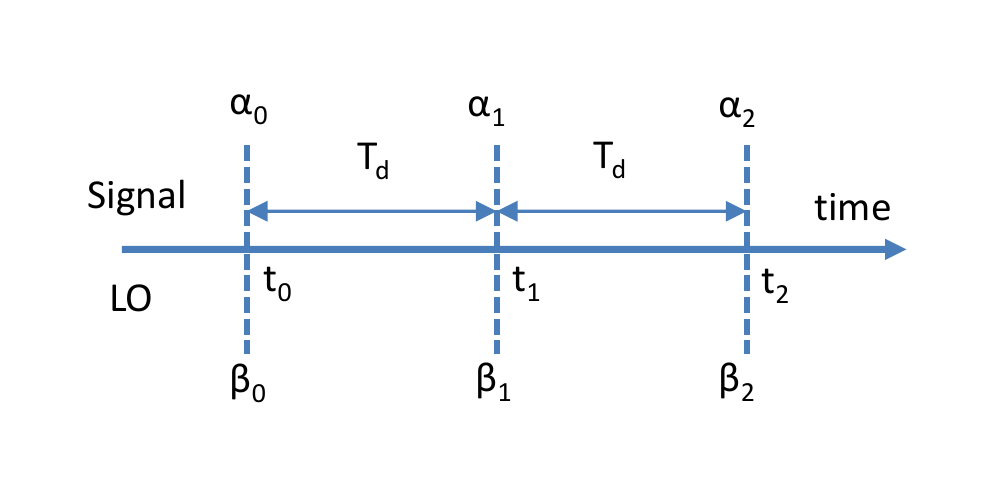}
	\captionsetup{justification=raggedright,
					singlelinecheck=false }
	\caption{Phase noise analysis. } 
	\label{fig:5}
\end{figure}

Assume that the phases of the signal laser and the LO laser at time $\left\lbrace t_0, t_1, t_2\right\rbrace $ are $\left\lbrace \alpha_0, \alpha_1, \alpha_2; \beta_0, \beta_1, \beta_2\right\rbrace $ correspondingly. The phase difference of the two lasers at the above times are given by 
\bes\label{eqa1} \phi_0=\beta_0-\alpha_0 \nonumber\\
\phi_1=\beta_1-\alpha_1 \nonumber\\
\phi_2=\beta_2-\alpha_2. \ees

The phases of the signal laser at different times are related by 
\bes\label{eqa2} \alpha_1=\alpha_0+2\pi f_S T_d+N_{S,1} \nonumber\\
\alpha_2=\alpha_1+2\pi f_S T_d+N_{S,2}, \ees
where $f_S$ is the central frequency of the signal laser. $N_{S,1}$ and $N_{S,2}$ are independent Gaussian noises with a mean of zero and a variance of $\langle(\Delta\theta_S(T_d))^2\rangle$.

Similarly, the phases of the LO laser are related by 
\bes\label{eqa3} \beta_1=\beta_0+2\pi f_L T_d+N_{L,1} \nonumber\\
\beta_2=\beta_1+2\pi f_L T_d+N_{L,2}, \ees
where $f_L$ is the central frequency of the LO laser. $N_{L,1}$ and $N_{L,2}$ are independent Gaussian noises with a mean of zero and a variance of $\langle(\Delta\theta_L(T_d))^2\rangle$.

We assume that $\phi_0$ and $\phi_2$ can be determined precisely by using strong reference pulses. From Eq.(6) and using Eqs.(A1-A3), phase $\phi_1$ can be estimated by
\bes\label{eqa4} \overline{\phi_1}=\dfrac{\phi_0+\phi_2}{2} \nonumber\\
=\phi_1+\dfrac{N_{S,1}+N_{L,2}-N_{S,2}-N_{L,1}}{2}. \ees

Since all the above noise terms in Eq. (A4) are independent with each other, it is easy to show the noise variance of the phase recovery scheme is given by
\bes\label{eqa5} \sigma_{\phi_1}=\langle(\overline{\phi_1}-\phi_1)^2\rangle \nonumber\\
=\dfrac{1}{2} \left\lbrace \langle(\Delta\theta_S(T_d))^2\rangle + \langle(\Delta\theta_L(T_d))^2\rangle  \right\rbrace. \ees
This is Eq.(11) in the main text.

We conduct experiments to determine the laser phase noise $\langle(\Delta\theta_S(T_d))^2\rangle$ and $\langle(\Delta\theta_L(T_d))^2\rangle$. The experimental setup is shown in Fig.7. The cw output of a laser is split into two beams by a symmetric fiber splitter. After the the two beams passing through two separate fiber links, the phase difference between the two beams are measured with a $90^{o}$ optical hybrid, two balanced photodetectors, and an oscilloscope.

\begin{figure}[t]
	\includegraphics[width=.45\textwidth]{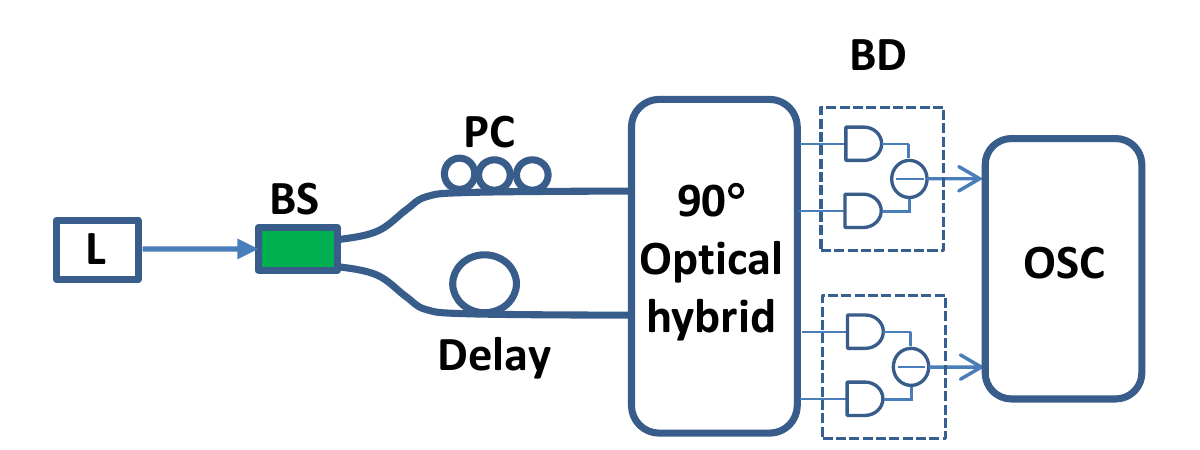}
	\captionsetup{justification=raggedright,
					singlelinecheck=false }
	\caption{Experimental setup for determining laser phase noise. L-laser; BS-fiber beam splitter; PC-polarization controller; BD-balanced photodetector; OSC-oscilloscope. } 
	\label{fig:7}
\end{figure}

Given the time delay difference between the two fiber links is $T_d$, we can determine the phase noise $\langle(\Delta\theta(T_d))^2\rangle$ of each laser directly. The phase noise of both the signal laser and the LO laser are measured at time delay $T_d=(5ns, 20ns, 25ns)$. The experimental results are shown in Fig.8. As expected from Eq.(9), the observed laser phase noise linearly depends on $T_d$. At $T_d=20ns$, the phase noise of the two laser has been determined to be $0.035\pm0.001$ and $0.044\pm0.001$.

\begin{figure}[t]
	\includegraphics[width=.5\textwidth]{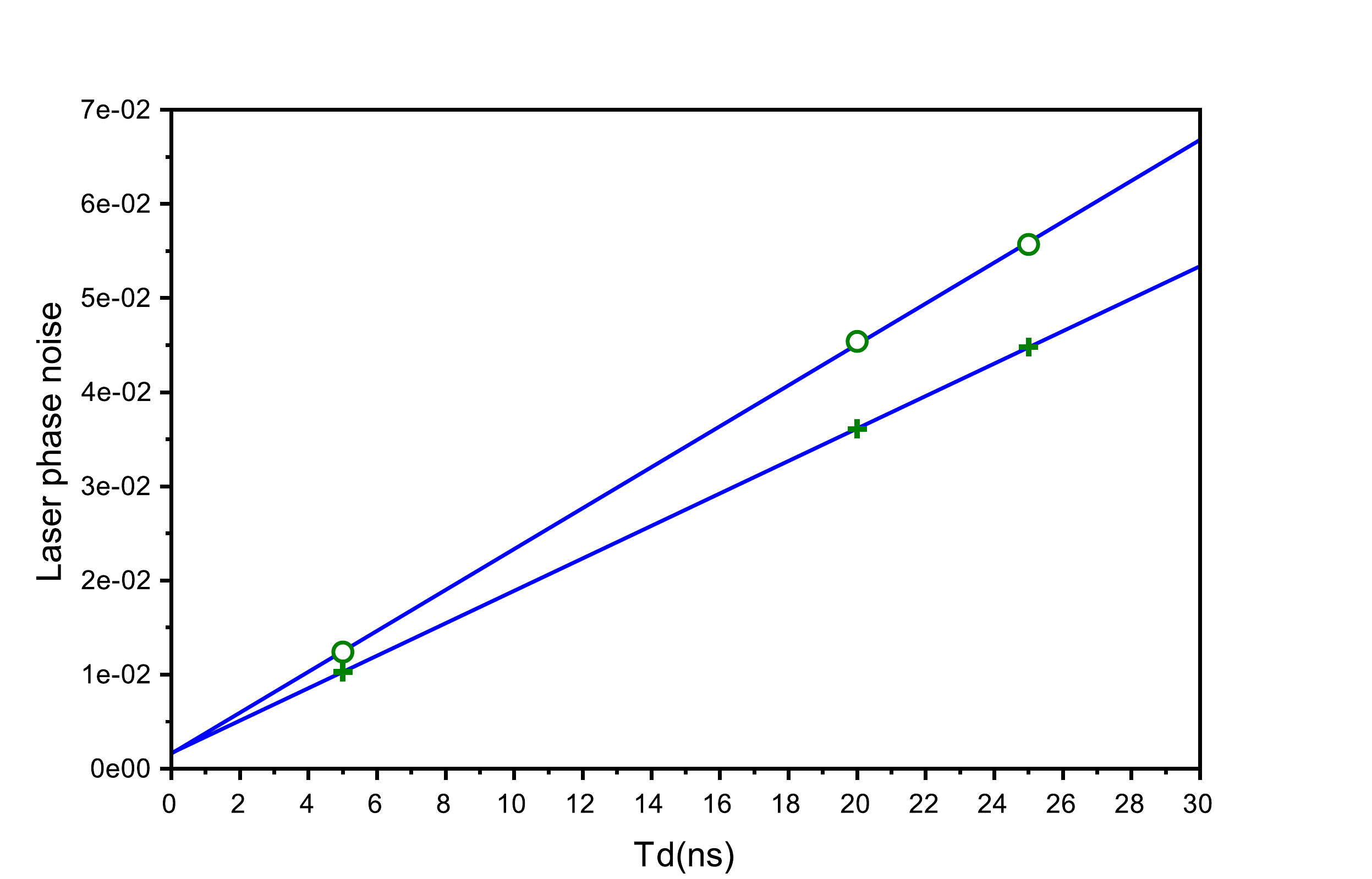}
	\captionsetup{justification=raggedright,
					singlelinecheck=false }
	\caption{Measured laser phase noise at different time delay $T_d$. `o'-LO laser; `+'-signal laser.} 
	\label{fig:8}
\end{figure}

\section{Simulation of secure key rate}

The security of one-way GMCS QKD has been well established. Here, our simulations are based on secure key rate formulas given in \cite{Fossier09}.

The secure key rate under the optimal collective attack, in the case of reverse reconciliation, is given by 
\bes\label{A1} R=fI_{AB}-\chi_{BE} \ees
where $I_{AB}$ is the Shannon mutual information shared between Alice and Bob; $f$ is the efficiency of the reconciliation algorithm; $\chi_{BE}$ is the Holevo bound of the information between Eve and Bob.
 
The mutual information between Alice and Bob is given by
\bes\label{A2} I_{AB}=log_2\dfrac{V+\chi_{tot}}{1+\chi_{tot}} \ees

The Holevo bound of the information between Eve and Bob is given by
\bes\label{A3} \chi_{BE}=\sum_{i=1}^2 G\left( \dfrac{\lambda_i-1}{2} \right)  - \sum_{i=3}^5 G\left( \dfrac{\lambda_i-1}{2}\right)  \ees
where $G(x)=(x+1)log_2(x+1)-xlog_2x$

\bes\label{A4} \lambda_{1,2}^2=\frac{1}{2} \left[ A\pm \sqrt{A^2-4B} \right] \ees
where
\bes\label{A5} A=V^2 (1-2T)+2T+T^2 (V+\chi_{line})^2 \ees
\bes\label{A6}B=T^2(V\chi_{line}+1)^2 \ees
 
\bes\label{A7} \lambda_{3,4}^2=\frac{1}{2} \left[ C\pm \sqrt{C^2-4D} \right] \ees
where
\begin{equation}
\begin{split}
C=\dfrac{1}{(T(V+\chi_{tot}))^2} [ A\chi_{het}^2+B+1+2\chi_{het} \\
( V\sqrt{B}+T(V+\chi_{line})) +2T(V^2-1)]
\end{split}
\end{equation}

\bes\label{A9}D=\left( \dfrac{V+\sqrt{B}\chi_{het}}{T(V+\chi_{tot})} \right) ^2 \ees 
    
\bes\label{A10} \lambda_5=1 \ees

\begin{figure}[t]
	\includegraphics[width=.5\textwidth]{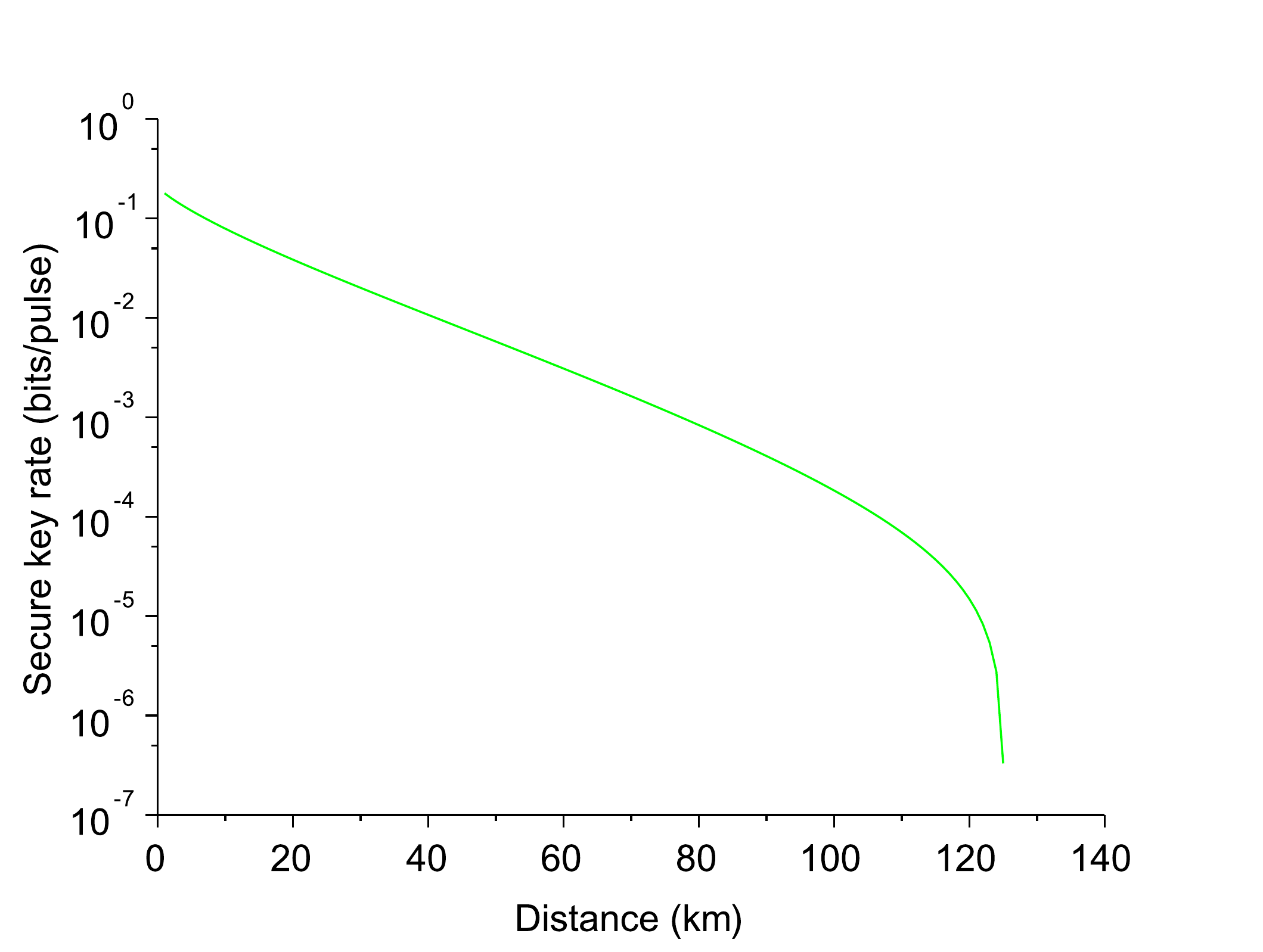}
	\captionsetup{justification=raggedright,
					singlelinecheck=false }
	\caption{Simulation results based on realistic parameters.} 
	\label{fig:9}
\end{figure}

System parameters in the above equations are defined as follows.
\begin{itemize}
\item[{\em (a)}] $V=V_A+1$, where $V_A$ is Alice's modulation variance.
\item[{\em (b)}] The total noise referred to the channel input $\chi_{tot}=\chi_{line}+\dfrac{\chi_{het}}{T}$, where $T$ is the channel transmittance. If we assume the quantum channel between Alice and Bob is optical fiber with an attenuation coefficient of $\alpha$, then the channel transmittance is given by $T=10^{\frac{-\alpha L}{10}}$, where $L$ is the fiber length.
\item[{\em (c)}] The total channel-added noise referred to the channel input $\chi_{line}=\frac{1}{T}-1+\varepsilon$, where $\varepsilon$ is the excess noise outside of Bob's system. We assume that $\varepsilon$ is mainly due to imperfection of the LO phase recovery scheme
\bes\label{A11} \varepsilon=V_A \sigma_\phi \ees
where $\sigma_\phi$ is the noise variance associated with the LO phase recovery scheme.
\item[{\em (d)}] The detection-added noise referred to Bob's input $\chi_{het}=[1+(1-\eta)+2\nu_{el}]/\eta$, where $\nu_{el}$ and $\eta$ are detector noise and detector efficiency, respectively.
\end{itemize}

We conduct numerical simulation using realistic parameters as summarized below: $\alpha=0.2$ dB/km, $\nu_{el}=0.1$, $\sigma_\phi=0.04$, $\eta=0.5$, $f=0.95$, and $V_A=1$. Fig.3 shows the simulation result in the asymptotic case. The simulation result shows that the proposed LO phase recovery scheme can be applied to achieve efficient QKD.

\begin{figure}[t!]
	\includegraphics[width=.45\textwidth]{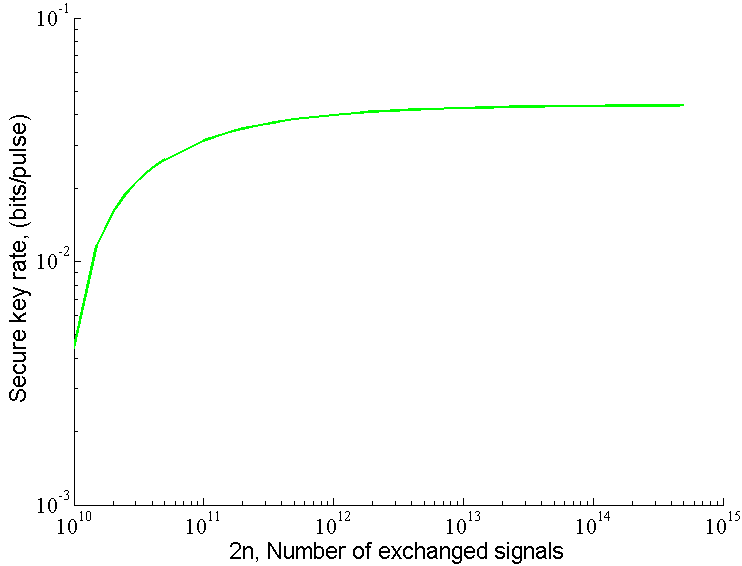}
	\captionsetup{justification=raggedright,
					singlelinecheck=false }
	\caption{Secure key rate simulation results for a finite number of pulses.} 
	\label{fig:10}
\end{figure}

Note that the secure key rate depicted in Fig.~\ref{fig:9} are obtained under the assumption of infinite number of pulses sent from Alice to Bob. However, experimentally one is always limited to a finite size data sample. To estimate the effect of finite data on the secure key rate we also conduct simulation using the most recent composable security proof~\cite{Leverrier15}. It can be shown (see Eq.(C13) in supplemental materials of~\cite{Leverrier15}) that the secure key rate under the optimal collective attack is,
\begin{eqnarray} \label{Eq:finiteR}
 R & = &  (1-\epsilon_{rob})(\beta I_{AB}-f(\Sigma^{max}_{a},\Sigma^{max}_{b},\Sigma^{min}_{c}) \nonumber \\ 
 & & - \frac{1}{2n}  [\Delta_{AEP}-\Delta_{ent}-2\log_{2}\frac{1}{2\bar{\epsilon}}])
\end{eqnarray}
where $I_{AB}$ is the Shannon mutual information shared between Alice and Bob given in Eq.(\ref{A2}); $\beta$ is the efficiency of the reconciliation algorithm; $f$ is the upper bound of the Holevo information $\chi_{BE}$ between Eve and Bob calculated in supplemental materials of~\cite{Leverrier15} (Eqs.(B2,C9-C11)); $\epsilon_{rob}$ is the protocol robustness parameter; $\Delta_{ent} = \log_{2}\frac{1}{\epsilon}-\sqrt{8n\log_{2}^2(4n)\log_{2}(1/\epsilon)}$ and $\Delta_{ent} = \sqrt{2n}[(d+1)^2 + 4(d+1)\log_{2}(2/\epsilon_{sm}^2)+2\log_{2}(2/\epsilon^2\epsilon_{sm})] - 4\epsilon_{sm}d/\epsilon$. For our simulations, following~\cite{Leverrier15}, we choose protocol parameters such that it is $\epsilon$-secure against collective attacks with $\epsilon=10^{-20}$ and $\epsilon_{cor}$-correct with $\epsilon_{cor}\le10^{-2}$ by setting $\epsilon_{sm}=\bar{\epsilon}=10^{-21}, \epsilon_{PE}=\epsilon_{cor}=\epsilon_{ent}=10^{-41}$. We also assume that the discretization parameter $d=5$ i.e. each measurement result is placed in one of five bins. Similarly to the asymptotic secure key rate simulations we set the physical parameters  $\alpha=0.2$ dB/km, $\sigma_\phi=0.04$, and $V_A=1$ and the reconciliation efficiency $\beta=0.95$. In Fig.~\ref{fig:10} we plot the simulated secure key rate as a function of the number of pulses transmitted for a fixed fiber length $L=10~km$ assuming perfect detectors
( $\nu_{el}=0$, $\eta=1$ ). The simulation results indicate that a usable secure key can be generated by sending $\approx 10^{11}$ pulses which is achievable with a CV-QKD system operated at tens of MHz rate.

\end{document}